\documentclass[aps,pra,twocolumn,superscriptaddress, floatfix]{revtex4-1}
\pdfoutput=1
\usepackage{amssymb}
\usepackage{amsmath}
\usepackage{bm}
\usepackage{graphicx}
\usepackage{textcomp}

\newcommand{\ket}[1]{\ensuremath{\left| #1 \right>}}
\newcommand{\bra}[1]{\ensuremath{\left< #1 \right|}}
\newcommand{\braket}[2]{\ensuremath{ \left< #1 \mid #2 \right>} }

\graphicspath{ {figures/} } 

\begin{document}
\title{Photon collection from a trapped ion--cavity system}

\author{J. D. Sterk}
\altaffiliation[Present Address: ]{Sandia National Laboratories, Albuquerque NM
87185}
\email{jdsterk@sandia.gov}
\author{L. Luo}
\altaffiliation[Present Address: ]{Department of Physics, Indiana
University--Purdue University Indianapolis, Indianapolis IN 46202}
\author{T. A. Manning}
\affiliation{Joint Quantum Institute, University of Maryland Department of
Physics and National Institute of Standards and Technology, College Park, MD
20742}
\author{P. Maunz} 
\affiliation{Fitzpatrick Institute for Photonics, Department of Electrical and
Computer Engineering, Duke University, Durham, NC 27708}
\author{C. Monroe}
\affiliation{Joint Quantum Institute, University of Maryland Department of
Physics and National Institute of Standards and Technology, College Park, MD
20742}

\date{\today}

\begin{abstract} 
  We present the design and implementation of a trapped ion cavity QED system. A
  single ytterbium ion is confined by a micron-scale ion trap inside a
  $2~\mathrm{mm}$ optical cavity.  The ion is coherently pumped by near resonant
  laser light while the cavity output is monitored as a function of pump
  intensity and cavity detuning.  We observe a Purcell enhancement of scattered
  light into the solid angle subtended by the optical cavity, as well as a
  three-peak structure arising from strongly driving the atom. This system can
  be integrated into existing atom--photon quantum network protocols and is a 
  pathway towards an efficient atom--photon quantum interface.
\end{abstract}

\maketitle

\section{Introduction}
Quantum networks rely upon an efficient interface between the quantum memories and
the photonic channels used for remote entanglement.  Trapped ions are a standard
platform to realize a quantum network due to their long coherence times and
the availability of high
fidelity quantum gate operations~\cite{blatt2008,duan2010}.  Current
implementations of trapped ion based quantum networks have demonstrated the
ability to entangle remote nodes, violate Bell's inequality, perform
teleportation, realize remote quantum gates, and generate private random
numbers~\cite{moehring2007b, matsukevich2008, olmschenk2009, maunz2009,
pironio2010}.  In these experiments, the interface between the ion and the
photonic channel depends upon a probabilistic process wherein the scattered photon is
collected by a microscope objective subtending only a small fraction of the
emission solid angle.  

There has been recent interest in integrating optical elements with an ion trap
system to improve the photon collection efficiency. Nearby optics
such as a fiber tip~\cite{vandevender2010}, reflective mirror~\cite{shu2009,
shu2010}, or Fresnel lens~\cite{streed2009,streed2011} can have larger numerical
apertures than common microscope objectives. Additionally, integrating
multi-scale optics---such as microfabricated mirrors---with ion traps may provide a
path towards scaling up a trapped ion network~\cite{noek2010,brady2011}.  
Although these methods increase the collection efficiency, they are still
inherently probabilistic in nature.

Coupling a trapped ion to an optical cavity can lead to a very high photon
collection efficiency. In principle, with a large coupling strength between an
ion and a cavity, the photon collection efficiency can approach
unity~\cite{law1997, mckeever2004}.  Since the cavity mode interacts coherently with the
atomic state, this process is reversible and can be used for generating quantum
networks~\cite{cirac1997}. Experiments with neutral atoms---where transition
frequencies are typically in the infrared---have demonstrated efficient atom--photon
interfaces as well as atom--photon and photon--photon
entanglement~\cite{mckeever2004, wilk2007, weber2009}. 

Efforts toward coupling trapped ions to optical cavities have used infrared
frequency transitions to a metastable D state for the optical
cavity~\cite{guthohrlein2001, mundt2002, keller2004, barros2009}.  At these
wavelengths, high finesse mirrors are available and strong coupling can be
achieved.  Single photons can be efficiently generated in this system using
techniques similar to neutral atom cavity QED experiments~\cite{keller2004,
barros2009}. However, these methods do not integrate directly with currently
demonstrated trapped ion quantum network protocols.

In this paper, we detail the design and fabrication of a trapped ion system
where a single ytterbium ion is coupled to a moderate-finesse optical cavity
resonant with the ultraviolet $S_{1/2}\leftrightarrow P_{1/2}$ transition at
$369.5~\mathrm{nm}$. Such a system could couple to individual hyperfine levels
of the ytterbium qubit and be integrated into existing atom--photon quantum
network protocols~\cite{luo2009}.  We trap a single $^{174}\mathrm{Yb}^{+}$ ion
inside the cavity and coherently pump it with a laser from the side of the
cavity while monitoring the cavity output. The photon scatter rate into the
solid angle subtended by the cavity mode in the outcoupling direction is
enhanced by a factor of $600$. Additionally, the spectral properties of the
atomic emission are observed as we detune the cavity from the atomic resonance.
At large pump strengths, an emergence of a three-peak structure from the
spectrum of emitted light indicates a Mollow triplet on a single atom level.

\section{Experimental System}
A single ytterbium ion is confined by a radiofrequency (RF) ion trap inside the mode of an
optical cavity resonant to the $S_{1/2} \leftrightarrow P_{1/2}$ transition at
$369.5~\mathrm{nm}$. To couple the ion to the optical cavity, we developed
a novel micron-scale ion trap that can be inserted into the cavity mode \emph{in
situ}. A nearby RF ground is an order of magnitude closer to the ion than the
dielectric mirrors to mitigate effects of dielectric charging~\cite{harlander2010}.

\subsection{Optical cavity design, fabrication, and test} 
The optical cavity consists of a pair of highly reflective concave mirrors from
Advanced Thin Films set up in a near-planar Fabry--P\'erot configuration. The
mirrors were initially $7.75~\mathrm{mm}$ in diameter and $4~\mathrm{mm}$ thick
when coated.  After coating, the mirrors were coned to a $2~\mathrm{mm}$
diameter reflective surface and $4~\mathrm{mm}$ outer diameter.  The radius of
curvature of the mirrors is $25~\mathrm{mm}$.

At ultraviolet frequencies, optical coating losses are two orders of magnitude
larger than infrared frequencies. The absorption and scattering losses for our
cavity initially was $\approx 400~\mathrm{ppm}$.  The mirrors form an asymmetric cavity with
an outcoupling mirror transmission of $T_{\mathrm{out}}\approx 1000~\mathrm{ppm}$, which is larger
than the incoupling mirror (set to $T_{\mathrm{in}}\approx 200~\mathrm{ppm}$).  Because the
ultraviolet light for Doppler cooling and coherently pumping the ion is derived
from a frequency doubled diode laser, the mirrors were additionally coated for
$739~\mathrm{nm}$ for cavity length stabilization (cf. Sec.~\ref{sec:setup}).

The free spectral range of the optical cavity was measured to be
$70.5~\mathrm{GHz}$ by scanning the $739~\mathrm{nm}$ laser across consecutive
transmission peaks.  This corresponds to a cavity length of $2.126~\mathrm{mm}$.
The full-width at half-maximum was measured by scanning the cavity length across
resonance and using the frequency difference between acousto-optic modulator
(AOM) orders as a frequency marker~\cite{hood2001}. The measured full-width at
half-maximum was $\kappa/\pi = 18.6~\mathrm{MHz}$, corresponding to a finesse of
$\mathfrak{F}=3790$, and an outcoupling efficiency of $0.6$. 

However, we noticed that the finesse of our cavity degraded after running the
experiment for several weeks. To our knowledge, several other groups have
noticed a similar effect in their ultraviolet cavities~\cite{colombePC,
bylinskiiPC}. A direct measurement of the cavity linewidth was made by driving
the cavity mode at a fixed laser frequency and scanning the cavity length across
resonance.  After attenuation of the output, photon counts are measured on a
photomultiplier tube (PMT).  The increased linewidth is $\kappa/\pi =
47.4~\mathrm{MHz}$, corresponding to a finesse of $\mathfrak{F} = 1490$ and
outcoupling efficiency of $0.24$.  The relevant cavity QED parameters for our
system are thus $(g,\kappa,\gamma)/2\pi = (3.92,23.7,19.6)~\mathrm{MHz}$, which
gives a single-atom cooperativity of $C = g^{2}/\kappa\gamma = 0.033$. Due to
the comparable strengths of the parameters, our system lies in the intermediate
regime of cavity QED~\cite{childs1994, kimble1994}.

\subsection{Ion trap for enhanced light collection}
\begin{figure}
  \begin{center}
	\includegraphics[width=0.9\columnwidth,keepaspectratio]{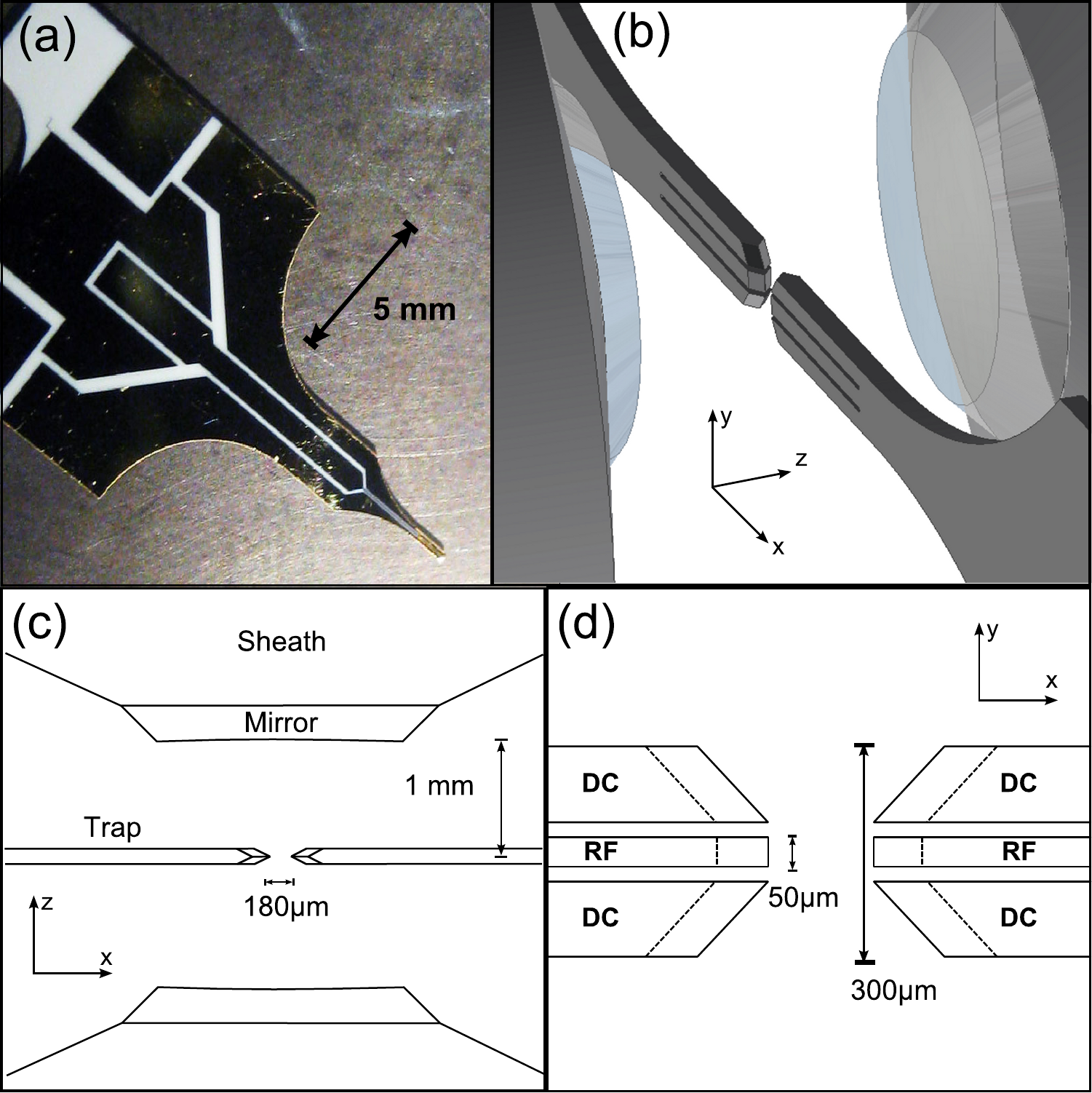}
  \end{center}
  \caption{Trap electrode and cavity geometry. 
  \emph{(a)} Photograph of a substrate ready for wiring. 
  \emph{(b)} Two identical substrates are inserted
  between the cavity mirrors such that the ion resides in the cavity mode.
  \emph{(c)} A top view of the ion trap inside the cavity. The substrates are
  separated by $180~\text{\textmu m}$ and are $1~\mathrm{mm}$ from the mirror.
  The mirrors are mounted in metal sheaths to provide compensation along the
  cavity direction.
  \emph{(d)}
  The tapered tip is machined to three
  individual tines. The outer tines are RF ground and provide compensation
  fields, while the center tine is RF high.}
  \label{fig:doublefork}
\end{figure}

Our cavity QED system requires an optically open ion trap where the ion can be
precisely placed inside the optical mode. The RF quadrupole trap is a modified version of an
earlier design, where the ion trap electrodes can be adjusted
independently~\cite{deslauriers2006a}.  The designed trap consists of two
identical laser machined alumina substrates with lithographically patterned
electrodes (Figure~\ref{fig:doublefork}).  Each substrate is mounted on a linear
positioner inside the vacuum chamber such that the position of the ion trap can
be placed inside the cavity mode \emph{in situ}.

Figure~\ref{fig:doublefork}a is a photograph of a finished substrate.  Each 
substrate is laser machined to a narrow finger approximately $300~\text{\textmu m}$
wide.  This finger is further machined into three individual tines, as
illustrated in Fig.~\ref{fig:doublefork}d.  The outer tines are approximately
$100~\text{\textmu m}$ wide and provide a nearby RF ground.  The center tine is
approximately $50~\text{\textmu m}$ wide and is at RF high.  The gaps between the tines
are about $25~\text{\textmu m}$, and the tips are beveled on both sides to minimize surface area visible to the ion. The back portion of the electrodes  provides enough
space for onboard RF filters that are wire bonded onto the substrate.  Gold is
evaporated onto the surface of the substrate up to a thickness of
$1~\text{\textmu m}$.  A particular lithographic process was developed to ensure gold was
coated around the entire substrate tip.

Stray electric fields can be compensated by applying static DC offset voltages
to the outer tines.  To compensate for stray fields along the cavity axis, we
apply a DC voltage on metal sheaths placed around the cavity mirrors as shown in
Figure~\ref{fig:doublefork}c.

To avoid diffractive losses in the cavity mode due to the presence of the ion
trap, the trap separation was chosen to be $180~\text{\textmu m}$, which is greater
than three times the mode diameter $2w_{0} = 50~\text{\textmu m}$.  At this
separation, the reduction in trap confinement compared to an ideal hyperbolic
electrode trap of the same characteristic size---known as the voltage
efficiency factor $\eta$---is $0.45$.  We apply $300~\mathrm{V}$ of RF at
$21.6~\mathrm{MHz}$,
and observe secular frequencies of $4~\mathrm{MHz}$.

The secular frequencies are measured by resonantly driving the secular motion of
the ion with a sinusoidal voltage on one of the outer tines.  While monitoring
the ion motion on a CCD camera, the frequency of the voltage is swept across the
motional resonances. From the orientation of the motion on the camera, we are able to
discriminate between motion along the cavity axis and along the trap axis.

\begin{figure}
  \begin{center}
	\includegraphics[width=\columnwidth,keepaspectratio]{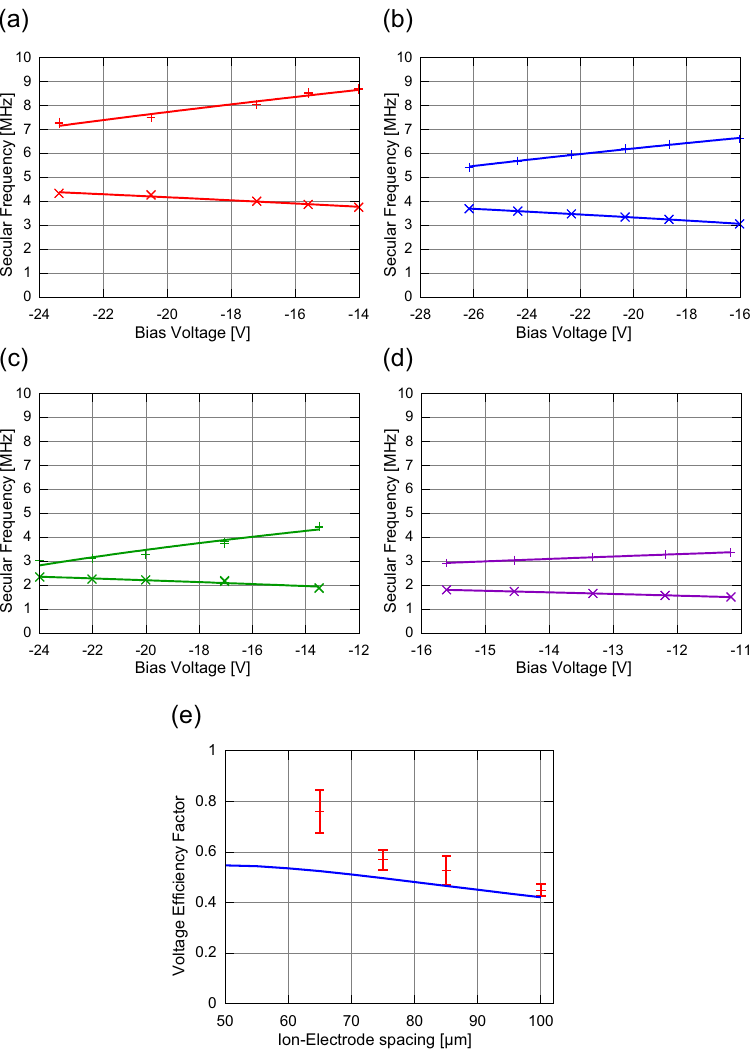}
  \end{center}
  \caption{Electrical characterization of our trap. The secular
  frequency of the trap is measured for various bias voltages $U_{0}$ and
  trap separations. Crosses ($+$) indicate frequencies along the trap axis,
  while ($\times$) indicate frequencies along the cavity axis. Solid lines
  indicate fits to equation~\ref{eqn:secFreq}. Electrode separations are: \emph{(a)} 
  $2x_{0}=130~\text{\textmu m}$, \emph{(b)} $150~\text{\textmu m}$. \emph{(c)}
  $170~\text{\textmu m}$. \emph{(d)} $200~\text{\textmu m}$. 
  \emph{(e)} The voltage efficiency factor from our measurements is compared to
  numerical simulations.
  }
  \label{fig:secFreq}
\end{figure}

We perform this measurement as a function of ion trap separation as well as a
DC bias voltage on the RF electrodes.  To lowest order, the secular frequencies of
the trap are given by
\begin{equation}
  \omega_{i} = \sqrt{ \frac{eU_{0}Q_{i}}{m} + \frac{1}{2}\left(
  \frac{e V_{0} Q_{i}}{m\Omega} \right)^{2}},
  \label{eqn:secFreq}
\end{equation}
where $Q_{i}$ is the quadrupole moment of the trap potential in the $i$-th
direction. Since the quadrupole moment is traceless, we note that $Q_{x}+Q_{y} +
Q_{z}=0$. Along the
electrode axis, the quadrupole moment is $Q_{x} = \eta/x_{0}^{2}$,
where $2x_{0}$ is the separation of the trap electrodes.  The voltage on the
RF electrode consists of a DC bias $U_{0}$ and RF voltage $V_{0}$ at frequency
$\Omega$.  Figures~\ref{fig:secFreq}a--d illustrates the measured secular
frequencies for various separations and bias voltages.  The solid lines are fits
to the data, providing good agreement with equation~\ref{eqn:secFreq}.  From the
data, we are able to extract the voltage efficiency factor for various
separations and compare it to a numerical simulation of the trap
(Figure~\ref{fig:secFreq}e).  

\subsection{Experimental Setup}  \label{sec:setup}
\begin{figure}
  \begin{center}
	\includegraphics{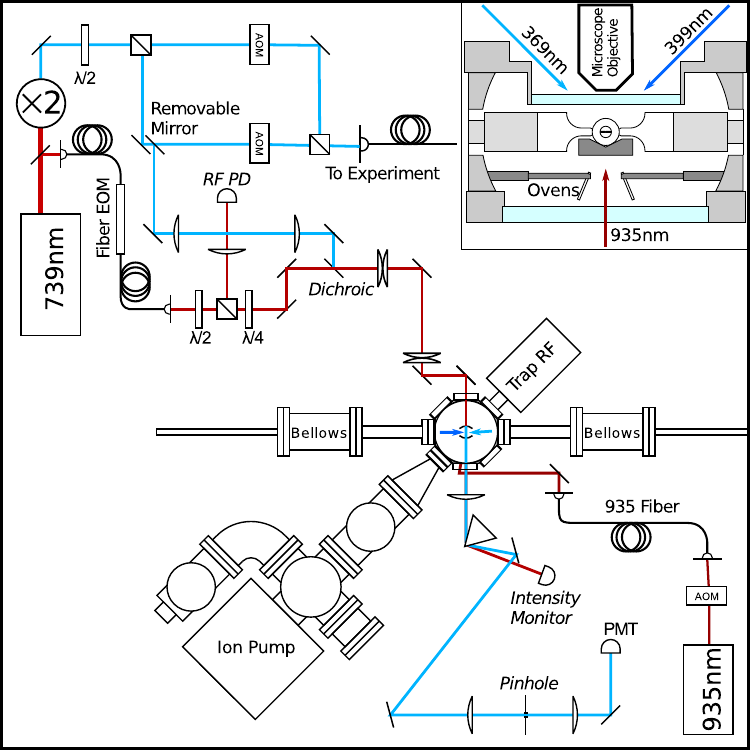}
  \end{center}
  \caption{The experimental system consists of a diode laser at
  $739~\mathrm{nm}$ that is frequency doubled to drive the atom. The two arms of
  the $369.5~\mathrm{nm}$ light are combined into a fiber that delivers the
  cooling and pump light to the ion. The cooling and pump are toggled during the
  experiment, allowing the pump to vary in intesity while keeping the cooling
  constant. The cavity is stabilized to the fundamental at $739~\mathrm{nm}$. A
  fiber electro-optic phase modulator (EOM) provides tunability of the cavity
  resonance as well as the necessary frequency offset to ensure
  co-resonance of the beams.  \emph{(Top Right)} A side view schematic along the
  cavity axis indicating the locations of the cooling/pump beam, the ionization
  beam, and the repump beam. The ion is imaged from above.} 
  \label{fig:setup}
\end{figure}

A diagram of the experimental apparatus is shown in Fig.~\ref{fig:setup}.  The ion
trap substrates are attached to individual linear positioners, allowing them to
be placed inside the cavity mode with micron-level precision. To coarsely align
the ion to the cavity mode, the cavity is repeatedly scanned over resonance with
$739~\mathrm{nm}$ light pumping the cavity. The transmission peaks are monitored
as the electrodes are inserted into the cavity mode. Loss of transmission
indicates a rough estimate of the position of the mode. Finer adjustment of the ion
position is made by
pumping the cavity with ultraviolet light and increasing the photon scatter from
the ion out
the side of the cavity.  The final method of improvement of ion--cavity coupling
is an iterative process where the fluorescence out the cavity is monitored and
maximized.

A thermal beam of ytterbium is produced by resistive heating of a stainless
steel tube in which Yb metal is packed. To minimize the probability of coating the cavity mirrors with
ytterbium, current is run through the ovens slightly above the threshold to
produce ytterbium atoms. Additionally, the thermal beam of atoms is
perpendicular to the cavity axis.  Ions are produced by a resonantly enhanced
two-photon transition to the continuum with a $398.9~\mathrm{nm}$ and
$369.5~\mathrm{nm}$ beam~\cite{olmschenk2007}.

The ions are cooled on the $S_{1/2} \leftrightarrow P_{1/2}$ transition of
$~^{174}\mathrm{Yb}^{+}$ at $369.5~\mathrm{nm}$ by a frequency doubled diode laser at
$739~\mathrm{nm}$. A $935.2~\mathrm{nm}$ beam is used to repump the ion out of a
low-lying $D_{3/2}$ level, which has a lifetime 
($52.7~\mathrm{ms}$) longer than the measurement time.
To measure the background light, the repump light is
turned off with an AOM.

The optical cavity is stabilized through a Pound--Drever--Hall locking technique
with $200~\text{\textmu W}$ of $739~\mathrm{nm}$ light at a frequency $\nu_{\mathrm{ir}}$.
Due to the Gouy phase shift and differences in the indicies of refraction of the
optical coating at $739~\mathrm{nm}$ and at $369.5~\mathrm{nm}$, the second
harmonic $2\nu_{\mathrm{ir}}$ is not resonant with the cavity. From the resonance
condition of the $q$-th longitudinal mode of a symmetric optical cavity of length $L$ and mirror radii of
curvature $\mathcal{R}$~\cite{siegman},
\begin{equation}
  \pi q = \frac{2\pi\nu_{q} L}{c} - \arccos \left( 1 - \frac{L}{\mathcal{R}}
  \right),
  \label{eqn:resonance}
\end{equation}
we find the ultraviolet resonace $\nu_{\mathrm{uv}}$ to
be shifted from the harmonic of the infrared resonance by an amount $\Delta f =
\nu_{\mathrm{ir}} - \frac{1}{2}\nu_{\mathrm{uv}}$ to be
\begin{equation}
 \Delta f
  = \nu_{\mathrm{ir}} \frac{L_{\mathrm{uv}} - L_{\mathrm{ir}}}{L_{\mathrm{uv}}} +
  \frac{c}{2\pi L_{\mathrm{uv}}} \left[ \phi_{\mathrm{ir}}
  -\frac{1}{2}\phi_{\mathrm{uv}} \right].
  \label{eqn:freq_shift}
\end{equation}
In equation~\ref{eqn:freq_shift}, we take the cavity lengths in the
ultraviolet and infrared to be $L_{\mathrm{uv}}$ and $L_{\mathrm{ir}}$, and the Gouy phase
$\phi_{\mathrm{ir(uv)}} = \arccos\left(1 - \frac{L_{\mathrm{ir(uv)}}}{\mathcal{R}_{\mathrm{ir(uv)}}} \right)$.
For our cavity, we find there to be a $2.3~\mathrm{GHz}$ frequency offset of the
infrared light to reach the resonance of the ytterbium ion, corresponding to a
length difference of approximately $12~\mathrm{nm}$.  The offset is provided by
a wide-bandwidth fiber EOM.  This modulator additionally gives us independent
control of the cavity resonance with respect to the laser frequency.

With the cavity locked, both ultraviolet light from the ion and infrared light
exit the cavity.  After an initial collimating lens, the output of the cavity is
sent through a prism to separate the colors (Figure~\ref{fig:setup}).  The infrared light is sent onto a
photodiode to monitor the cavity transmission, while the ultraviolet beam is
spatially filtered and directed onto a PMT. The overall efficiency of our
detection system (including detector quantum efficiency) is $4\%$.

\section{Experimental Results}
The experimental procedure is as follows.  The ion is Doppler cooled for
$20~\mathrm{ms}$ with a weak cooling beam.  Next, a strong pump beam is turned on
for $2~\mathrm{ms}$ and photon counts out of the cavity are recorded.  For half
the detection time, the $935~\mathrm{nm}$ repump light is off, providing a measurement of
the background light for subtraction. We average over $40$ measurement cycles
before changing the cavity frequency. In this manner, a lineshape is built up
for a set pump intensity. The strong pump is detuned from atomic resonance by
$10~\mathrm{MHz}$ to avoid heating of the ion during the measurement.

The power of the coherent pump is calibrated by ion fluorescence from the side of the
cavity.  The ion fluorescence is collected by a microscope objective and imaged
onto a PMT. By measuring the scatter rate out the side of the cavity for various
input powers, the intensity at the ion versus the input power can be determined
in terms of the saturation intensity.

When the cavity is resonant, we observe up to $8000$ photon counts per second on
the PMT. Our PMT has a quantum efficiency of $19\%$ and the prism has a
tranmission of $23.5\%$. Given these efficiencies in the detection path, our
measured value corresponds to $200,000$ photons emerging from the cavity per
second. From our estimated outcoupling efficiency of $0.24$, we
estimate the cavity collects $800,000$ photons per second. Table~\ref{tab:effic}
summarizes the efficiencies and backs out the photon scatter rate into the
cavity. The photon emission rate agrees with the estimate that the emission rate
is given by $p_{coll}\Gamma_{sc}$, where $p_{coll}$ is the collection probability
given by~\cite{luo2009}
\begin{equation}
  p_{coll} = \frac{T_{\mathrm{out}}}{\mathcal{L}}
  \left(\frac{2\kappa}{2\kappa+\gamma} \right)
  \left(\frac{2C_{\mathrm{eff}}}{1+2C_{\mathrm{eff}}}\right)
  \label{eqn:pcoll}
\end{equation}
and $\Gamma_{sc}$ is the photon scatter rate.
Here, $T_{\mathrm{out}}/\mathcal{L}$ is the outcoupling efficiency while
$2\kappa/(2\kappa+\gamma)$ is the ratio of the rate at which the photon leaves
the cavity to the total rate at which the ion--cavity system loses photons. The
third factor, $2C_{\mathrm{eff}}/(1+2C_{\mathrm{eff}})$, is the Purcell enhancement with an
effective cooperativity $C_{\mathrm{eff}}$ for a reduced coherent coupling rate due to
averaging of the atomic motion across the cavity standing wave.

\begin{table}
  \centering
  \begin{tabular}{ c | c | c |}
		&	Efficiency	& Count rate\\
	\hline
	Detected			&					& $8000~\mathrm{s}^{-1}$ \\
	Before PMT			&		$0.19$		&	$42,000~\mathrm{s}^{-1}$ \\
	Before Prism		&		$0.235$		&	$180,000~\mathrm{s}^{-1}$ \\
	Before vacuum window 	&	$0.9$		&	$200,000~\mathrm{s}^{-1}$ \\
	Outcoupling efficiency	&	$0.24$	&	$\approx800,000~\mathrm{s}^{-1}$ \\
  \end{tabular}
  \caption{List of efficiencies and the effective photon scatter rates before
  elements. There are approximately $800,000$ photons scattered into the cavity
  per second, where only $200,000$ photons/sec emerge from the cavity.}
  \label{tab:effic}
\end{table}

To compute the enhancement in photon collection efficiency, we compare our
result to isotropic scattering of photons from a single ytterbium ion. We define
the enhancement to be the ratio of photons emerging from the cavity to the
isotropic scatter rate into the solid angle of the cavity mode in outcoupling
direction.  The solid angle subtended by the cavity mode is given by
\begin{equation}
  \Delta \Omega_{\mathrm{cav}} = \frac{2\lambda^{2}}{\pi w_{0}^{2}} =
  1.465\times10^{-4}~\mathrm{sr}
  \label{eqn:solidangle}
\end{equation}
which is twice the solid angle subtended by the mode in the outcoupling
direction.  We calculate the photon scatter rate for isotropic scattering into
this solid angle for the parameters where we observe maximal counts on our PMT.
We estimate that without enhancement we would observe $300$ photons per second
scattered into the outcoupling direction of the cavity mode. 

This yields a spontaneous emission enhancement factor of $600$ compared to the
free-space value into the same solid angle of the cavity mode in the outcoupling
direction. Other cavity QED experiments in the intermediate regime of cavity QED
have reported an enhancement of $18.5$ of spontaneous emission into an undriven
cavity mode. This was achieved by delivering cold atoms from a magneto-optical
trap into the cavity~\cite{terraciano2007}. Due to the strong confinement of our
atom, our system achieves a much higher enhancement of
spontaneous emission than any other experiment in the intermediate regime.

\begin{figure}
  \begin{center}
	\includegraphics[width=\columnwidth,keepaspectratio]{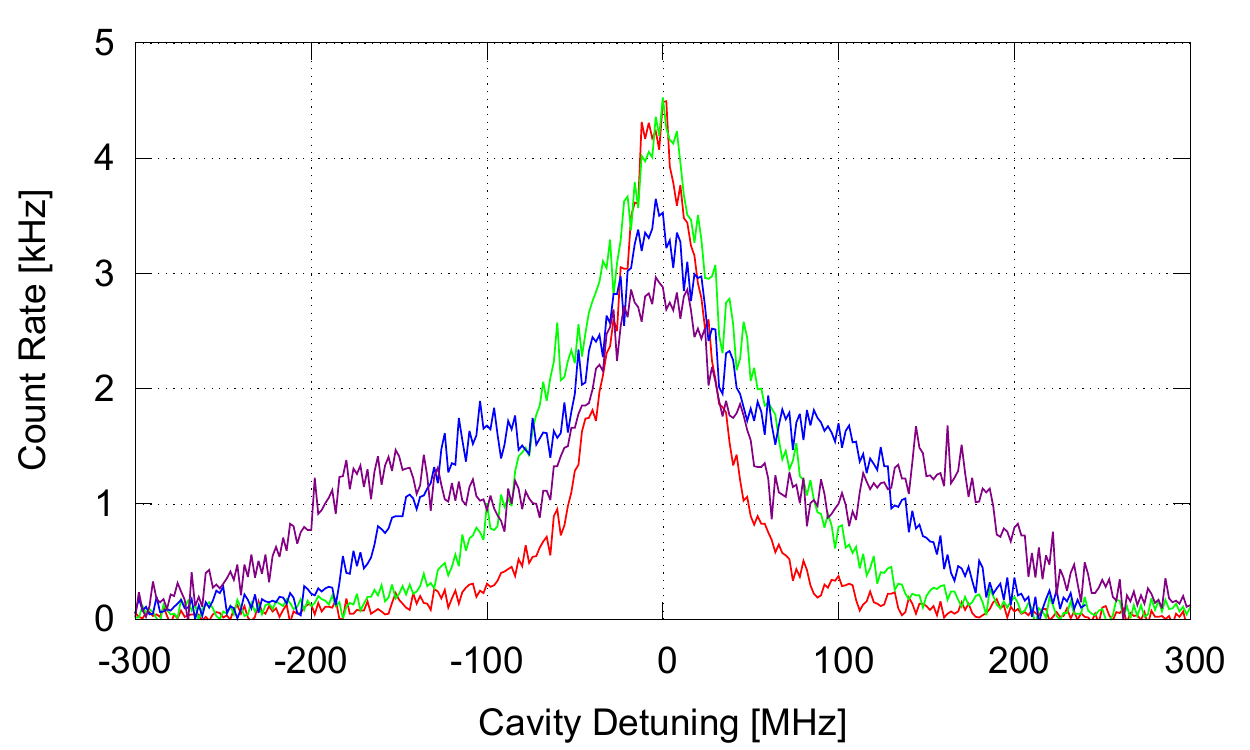}
  \end{center}
  \caption{Steady-state photon count rate from the cavity versus cavity detuning
  for several drive intensities. At intensities large compared to the saturation
  intensity, we observe the onset of a triplet structure. Red:
  $I/I_{\mathrm{sat}} = 2$, Green: $I/I_{\mathrm{sat}} = 50$, Blue:
  $I/I_{\mathrm{sat}} = 150$, Purple: $I/I_{\mathrm{sat}} = 600$.
  }
  \label{fig:composite}
\end{figure}

In our experiment, we measure the photon counts from the cavity as a function
of the cavity detuning $\delta_{c} = \omega_{c} - \omega_{L}$. The fiber EOM
offset allows us to set the cavity resonance independently of the pump
frequency We set the atom--laser detuning $\delta_{0} = \omega_{0} - \omega_{L}$
to be $10~\mathrm{MHz}$ below resonance, and measure the count rate versus the
cavity detuning.  Figure~\ref{fig:composite} illustrates the lineshapes
observed for the coherently driven atom for various pumping strengths
($I/I_{\mathrm{sat}} = 2, 50, 150, 600$).  At low
intensities, we observe a Lorentzian lineshape consistent with a cavity
broadened emission line.  For strong pump intensities
we observe the emergence of a three-peak structure characteristic of
the Mollow triplet~\cite{mollow1969}.  

\begin{figure}
  \begin{center}
	\includegraphics{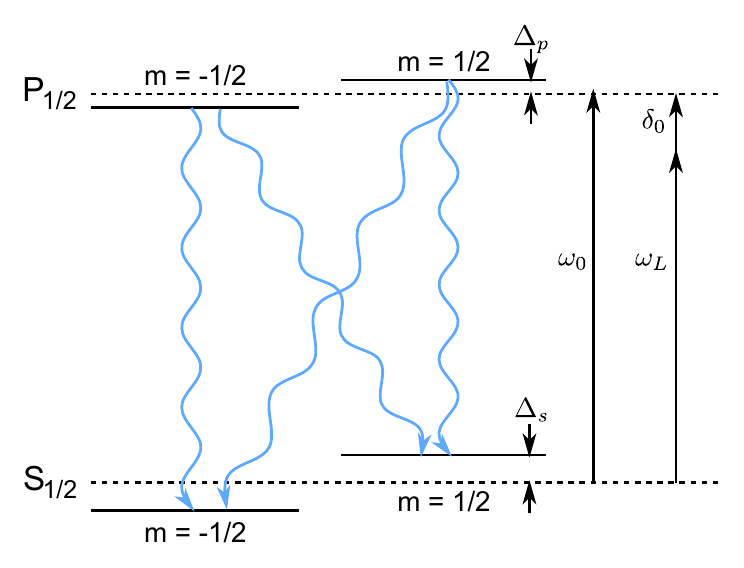}
  \end{center}
  \caption{Relevant atomic energy levels for the $~^{174}\mathrm{Yb}^{+}$ ion. The Zeeman
  levels are shifted by $\pm\hbar \Delta_{s,p}$. The pump light is detuned from the
  bare resonance by $\delta_{0}$, and the cavity detuning is $\delta_{c} =
  \omega_{c} - \omega_{L}$.
  }
  \label{fig:energylevels}
\end{figure}

We attribute this three-peak structure to a cavity-broadened fluorescence
spectrum of the ion. However, the fluoresence spectrum of a two level atom
cannot fully describe our data. Instead, the Zeeman levels with various driving
strengths must be taken into account.
The 174 isotope of ytterbium has zero nuclear spin, and
therefore has no hyperfine structure. The $S_{1/2}$ and $P_{1/2}$ manifolds each
consist of two nearly degenerate Zeeman states, split by a weak magnetic field
perpendicular to the cavity axis (Figure~\ref{fig:energylevels}).  

We label the four levels of interest as $\{\ket{S_{1/2},\pm},
\ket{P_{1/2},\pm}\}$, where $S_{1/2} (P_{1/2})$ corresponds to the ground
(excited) state
manifold. In a frame rotating at the classical drive frequency $\omega_{L}$, the
atomic Hamiltonian is
\begin{equation}
  H_{\mathrm{a}} = \hbar \sum_{\ell, m} \Delta_{\ell}
  \ket{\ell,m}\bra{\ell,m} + \hbar \delta_{0} \sum_{m} \ket{e,m}\bra{e,m}
  \label{eqn:H_atom}
\end{equation}
where $\pm\Delta_{s(p)}$ is the Zeeman level shift of $\ket{S_{1/2}(P_{1/2}),\pm}$. The
second term is the energy of the excited state manifold in the rotated frame, where $\delta_{0} = \omega_{0} - \omega_{L}$.
 The classical driving field, $\mathbf{E} = \frac{1}{2}
E_{0} \bm{\epsilon} e^{-i\omega_{L}t} + c.c.$, can drive all three types of transitions
in the ion ($\Delta m = \pm1,0$), depending on the orientation of the polarization vector
$\bm{\epsilon}$ with respect to the
magnetic field. The interaction Hamiltonian with the classical field can be
written as
\begin{equation}
  H_{\mathrm{d}} = -\frac{\hbar\Omega}{2} \left[ \hat{\mathbf{A}} \cdot \bm{\epsilon} +
  \hat{\mathbf{A}}^{\dagger}\cdot\bm{\epsilon} \right]
  \label{eqn:H_drive}
\end{equation}
where $\Omega = \mu E_{0}/\hbar = \gamma \sqrt{I/2I_{\mathrm{sat}}}$ is the Rabi
frequency for the $S_{1/2} \leftrightarrow P_{1/2}$ transition, and
the lowering operator is
\begin{equation}
  \hat{\mathbf{A}} = \sum_{q,m,m'}\braket{S_{1/2},m;1,q}{P_{1/2},m'} \ket{S_{1/2},m}
  \bra{P_{1/2},m'} \mathbf{e}_{q}
\end{equation}
with $\mathbf{e}_{q}$ being the spherical basis vector. The saturation intensity
for the $S_{1/2} \leftrightarrow P_{1/2}$ transition is $I_{\mathrm{sat}} =
\hbar \omega_{0}^{3} \gamma/12\pi c^{2} = 50.7~\mathrm{mW}/\mathrm{cm}^{2}$.
The
vector component $\hat{A}_{q}$ describes the transition between the magnetic
sublevels $m$ and $m-q$ and is proportional to the Clebsch--Gordan coefficient
of that transition~\cite{boozerThesis}.

Additionally, we consider the two degenerate polarization modes of the cavity at
a frequency $\omega_{c}$ detuned from the drive frequency by $\delta_{c} =
\omega_{c} - \omega_{L}$. The Hamiltonian for the two cavity modes in the
rotating frame is
\begin{equation}
  H_{c} = \hbar\delta_{c} \sum_{p=H,V} \hat{a}_{p}^{\dagger} \hat{a}_{p}.
  \label{eqn:H_cav} 
\end{equation}
The
Jaynes--Cummings cavity interaction consists of the coupling of the three atomic
transitions to the two cavity modes, and is given by
\begin{equation} 
  H_{jc} = i \hbar g \sum_{p=H,V} \left[
  \hat{a}_{p}^{\dagger} (\hat{\mathbf{A}}\cdot \mathbf{e}_{p}^{*})
  -
   (\hat{\mathbf{A}}^{\dagger}\cdot \mathbf{e}_{p})\hat{a}_{p} 
  \right]
  \label{eqn:H_jc}
\end{equation}

\begin{figure}
  \begin{center}
	\includegraphics[width=\columnwidth,keepaspectratio]{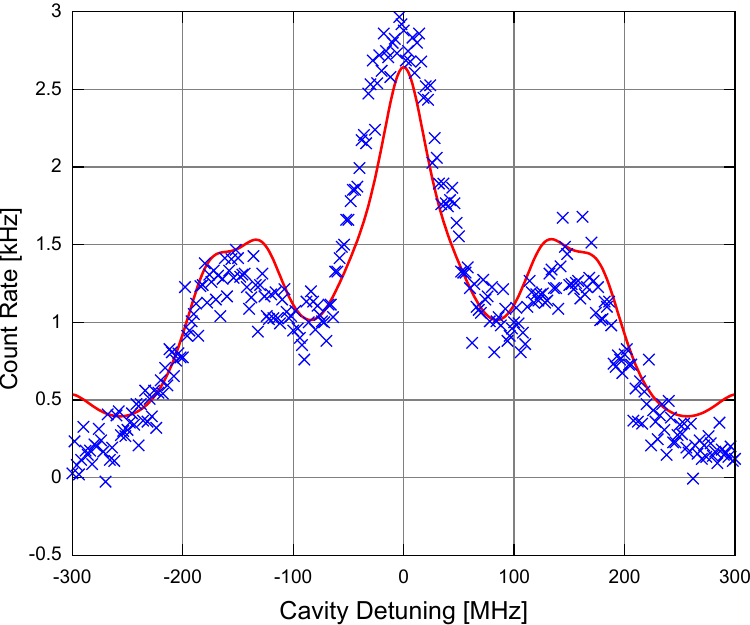}
  \end{center}
  \caption{Typical fit to our data. The solid curve is a numerical calculation
  of the steady-state intracavity photon number of our cavity versus cavity
  detuning $\delta_{c}/2\pi$. Taking into account the Zeeman levels in
  $~^{174}\mathrm{Yb}^{+}$,
  we reach qualitative agreement with the observed data.
  }
  \label{fig:compare}
\end{figure}

The steady-state photon count rate is given by the steady-state intracavity
photon number for both polarizations and the cavity decay rate: $2\kappa\left[ \langle n_{H}
\rangle_{ss} + \langle n_{V} \rangle_{ss} \right]$. To compute the intracavity
photon number, we numerically solve the master equation 
\begin{align}
  \dot{\rho} &=  -i [\rho,H_{\mathrm{a}} + H_{\mathrm{d}} + H_{c} + H_{jc}]  \\ \nonumber
  &+ \gamma \sum_{q} \left[ \hat{A}_{q} \rho
  \hat{A}_{q}^{\dagger} - \frac{1}{2} \left( \hat{A}_{q}^{\dagger} \hat{A}_{q}
  \rho + \rho \hat{A}_{q}^{\dagger} \hat{A}_{q} \right) \right] \\ \nonumber
  &+ 2\kappa \sum_{p}
  \left[ \hat{a}_{p} \rho \hat{a}_{p}^{\dagger} -  \frac{1}{2}\left(
  \hat{a}_{p}^{\dagger}\hat{a}_{p}\rho + \rho \hat{a}_{p}^{\dagger}\hat{a}_{p}
  \right) \right]
  \label{eqn:master}
\end{align}
in steady state using the
Quantum Optics Toolbox for \textsc{Matlab}~\cite{tan1999} with our experimental
values. The numerical
calculation is performed for various detunings of the cavity,
and parameters of the classical beam (intensity, orientation relative the
magnetic field, and polarizaiton). We fit the data to these simulations and
achieve qualitative agreement. A typical fit curve is illustrated in
Figure~\ref{fig:compare}, where the parameters for the classical beam are $I =
600I_{\mathrm{sat}}$, oriented $45^{\circ}$ from the magnetic field with linear
polarization tilted $35^{\circ}$ from the cavity axis.

Previously, strongly-driven atoms in a cavity QED system have been studied by
passing an atomic beam into an optical cavity, where the Mollow triplet has be
observed through the cavity output~\cite{zhu1988}. Extending such experiments to
a single trapped atom can achieve significant vacuum field dressed-state pumping
and steady-state inversion~\cite{lewenstein1988, savage1988, hughes2011}.
However, such experiments have never been done in trapped neutral atom cavity
QED experiments due to the relatively weak confinement compared to the atomic
recoil. The strong confinement of our trapped ion allows us to observe the
Mollow triplet transition coupled to a cavity at the single atom level for the
first time. These results open the way towards studying cavity QED physics with
trapped atoms in a strongly-driven regime. Superconducting qubits in a circuit
QED system have recently  been used to study similar physics~\cite{baur2009}.

\section{Conclusion}
We have designed and fabricated a trapped ion cavity QED system where we observe
a Purcell enhancement of scattered light into the cavity mode. The scatter rate
is two orders of magnitude larger than the scatter rate into the solid angle
subtended by the optical cavity.  Additionally, we have investigated the photon
count rate from the cavity as a function of the cavity detuning and the pump
strength. We have found at high intensities, the steady-state count rate
exhibits a three-peak structure characteristic of a Mollow triplet. Numerical
models of a $~^{174}\mathrm{Yb}^{+}$ ion coupled to an optical cavity yield
qualitative agreement with our data.

Additionally, we observe a degradation of the cavity finesse at ultraviolet
frequencies. This unexpected effect is possibly a materials related issue and
requires more study. Even at a distance of one millimeter from the ion, the
presence of the mirror had a noticable influence on the ion position, as
evidenced by a dynamic and variable stray electric field. This
necessitates better shielding and possibly a cavity in
a near-concentric configuration which would allow a larger mirror spacing while
maintaining the small mode volume required for adequate atom--cavity coupling.

Nevertheless, the enhancement of scattered light into the cavity mode
demonstrates the feasibility of increasing the photon collection efficiency
for quantum networks. Such an ion--cavity system can be readily integrated into
current protocols for trapped-ion quantum networks, providing a pathway towards a
practical, scalable quantum network.
For example, 
ion--photon entanglement with the polarization degree of
freedom~\cite{matsukevich2008,luo2009} is most amenable to an optical cavity.
An optical cavity can be locked to the $\ket{S_{1/2},F=1} \leftrightarrow
\ket{P_{1/2}, F=1}$
transition of $~^{171}\mathrm{Yb}^{+}$ with the quantization axis along the cavity
mode. A weak $\pi$-polarized probe on an ion initialized to the
$\ket{S_{1/2},F=0,m_{F}=0}$ state can excite the atom to the
$\ket{P_{1/2},F=1,m_{F}=0}$ level, which
is coupled to the $\ket{S_{1/2},F=1,m=\pm1}$ ground states through the two
polarization modes of the cavity. Such a procedure can produce ion--photon
entangled pairs that are useful for ion--photon quantum networks,
loophole-free tests of Bell inequalities, and the generation of cluster states.

\begin{acknowledgments}
  The authors would like to thank Luis A. Orozco and Howard Carmichael for
  providing useful insight into cavity QED theory as well as discussions of
  experimental techniques and system modeling.
 This work is supported by the US Army Research Office (ARO) with 
funds from the IARPA MQCO Program and the MURI program on Quantum 
Optical Circuits of Hybrid Quantum Memories, NSF Physics at the 
Information Frontier Program, and the NSF Physics Frontier Center at JQI.
\end{acknowledgments}

\bibstyle{apsrev}
\bibliography{bibfile}

\end{document}